\documentclass{elsart}

\usepackage{graphicx}

\usepackage{amssymb}

\begin{document}

\begin{frontmatter}

\title{Probabilistic Breakdown Phenomenon at  On-Ramp Bottlenecks in Three-Phase Traffic Theory}

\author{Boris S. Kerner}

\address{ 
DaimlerChrysler AG, REI/VF, HPC: G021, 71059 Sindelfingen, Germany 
}
and
\author{Sergey L. Klenov} 

\address{ 
Moscow Institute of Physics and Technology, Department of Physics, 141700 Dolgoprudny,
Moscow Region, Russia
}



\maketitle

\begin{abstract}
 A nucleation model for the    breakdown phenomenon in an initial non-homogeneous free traffic flow 
 that occurs at an on-ramp bottleneck is presented. This model
  is in the context of three-phase traffic theory. In this theory, the breakdown phenomenon is associated with
  a first-order phase transition from  the  $\lq\lq$free flow"   phase to the $\lq\lq$synchronized flow"   phase.
  In contrast with many other nucleation models for phase transitions in different
 system of statistical physics in which random precluster emergence from fluctuations
 in an initial homogeneous system foregoes  subsequent cluster evolution towards  a critical cluster (critical nuclei),
  random precluster occurrence in free flow at the bottleneck is not necessary for traffic breakdown. 
 In the model, the breakdown phenomenon can
 also occur if there were no fluctuations in free flow. This is because there is a permanent and motionless non-homogeneity that
 can be considered a deterministic vehicle cluster localized in a neighborhood of the bottleneck. The presented nucleation model and  
 a nucleation rate of traffic breakdown
 that follows from the model exhibit qualitatively different features in comparison with previous results. 
 In the nucleation model,  traffic breakdown nucleation occurs through a   random increase in vehicle number
   within  the  deterministic vehicle cluster, if the  amplitude of the resulting random vehicle cluster 
 exceeds some critical amplitude. 
   The mean time delay and the associated   nucleation rate of traffic breakdown at the bottleneck are   found and investigated. The  nucleation rate
   of traffic breakdown as a function of the flow rates to the on-ramp and upstream of the bottleneck is studied. 
  Boundaries for traffic breakdown in the diagram of congested patterns at the bottleneck are found.
    These boundaries are qualitatively correlated with numerical results of simulation of microscopic traffic flow models in the context
    of three-phase traffic theory.
\end{abstract}
\end{frontmatter}

\section{Introduction}
\label{Introduction}

Empirical observations of freeway traffic made in various countries show that  the onset of congestion
in an initial free flow is associated with
 an abrupt decrease in vehicle speed. This traffic breakdown called the $\lq\lq$breakdown phenomenon"  occurs 
 mostly at freeway bottlenecks, in particular  on-ramp bottlenecks. The traffic breakdown is
 accompanied by a hysteresis effect (see references in the reviews~\cite{Hall1992A,Banks2002A}, the book~\cite{KernerBook},
 and the conference proceedings~\cite{Lesort,Ceder,Taylor}). The breakdown phenomenon has a probabilistic nature~\cite{Elefteriadou1995A,Persaud1998B,Lorenz2000A}:
 At the same on-ramp bottleneck, traffic breakdown is observed at different flow rates in different realizations (days).
 The probability of the breakdown is a strong increasing function of  flow rate downstream 
 of the bottleneck~\cite{Persaud1998B,Lorenz2000A}.

Most microscopic, macroscopic, probabilistic, and other models of freeway traffic explain the onset of congestion in free flow
by  moving jam emergence (see, e.g.~\cite{KK1994,Bando,Schreck,Mahnke1997,Mahnke1999,Kuehne2002}
and references in the reviews~\cite{Sch,Helbing2001,Nagatani_R,Nagel2003A},
 as well as the conference proceedings~\cite{SW1,SW2,SW3,SW4,SW5}). In particular,  in models of
   an on-ramp bottleneck moving jam(s) occurs spontaneously in free flow at the bottleneck
when the flow rate upstream of the bottleneck is high enough
and the flow rate to the on-ramp increases gradually beginning from zero~\cite{Helbing2001,Nagatani_R,KKS1995,Helbing1999A,Lee1999A,Helbing2000}.
However, the fundamental model
 result that the onset of 
 congestion in free flow on a homogeneous road  
 and at freeway bottlenecks  
is associated
with spontaneous moving jam emergence~\cite{KK1994,Mahnke1997,Mahnke1999,Kuehne2002,KKS1995,Helbing1999A,Lee1999A,Helbing2000,Sch,Helbing2001,Nagatani_R,Nagel2003A}
 is in a serious conflict 
 with empirical evidence~\cite{Kerner1998B,Kerner2002B,KernerBook}. 
 
Consequently, in 1996--1999 Kerner introduced three-phase traffic theory 
(see~\cite{KernerBook} for a review). In this theory, there are three traffic phases:
free flow, synchronized flow, and wide moving jams. In accordance with   
  empirical investigations of phase transitions, in this theory
 moving jams do {\it not}
emerge spontaneously in   free flow. Rather than moving jam emergence, a phase transition from   free flow   to
 synchronized flow   (F$\rightarrow$S transition for short)
governs the onset of congestion in free flow~\cite{Kerner1998B,Kerner2002B}. A first-order F$\rightarrow$S transition postulated in three-phase traffic theory~\cite{Kerner1998B}
discloses the nature of the    breakdown phenomenon at freeway bottlenecks found in empirical observations~\cite{Hall1992A,Banks2002A}. In other words,
the terms $\lq\lq$F$\rightarrow$S transition", $\lq\lq$breakdown phenomenon", $\lq\lq$speed breakdown",
and $\lq\lq$traffic breakdown" are synonyms. 
 The first microscopic models in the context of three-phase traffic theory   
 are stochastic models~\cite{KKl,KKW}. These models exhibit phase transitions
  as well as all types of congested patterns found
  in empirical observations~\cite{KKl,KKW,KKl2003A,KKl2004AA,KernerBook}.
 Recently, some new  microscopic models based on three-phase traffic theory have been 
 developed~\cite{Davis2003B,Lee_Sch2004A,Jiang2004A}, which can show some congested pattern features 
found  earlier
in~\cite{KKl,KKW}. 

One of the important methods for a study of phase transitions in non-linear distributed multiple-particle systems is
a probabilistic theory based on an analysis of a master equation (e.g.,~\cite{Gardiner,vanKampen,MahnkeRev}). 
First probabilistic theories for traffic flow based on a master equation for a random vehicle cluster have been introduced by Mahnke et al.~\cite{Mahnke1997,Mahnke1999} and further developed by K\"uhne et al.~\cite{Kuehne2002,Kuehne2004} (see the recent review
by Mahnke et al.~\cite{MahnkeRev}).
As usual for other metastable systems of statistical physics, for an initial metastable traffic flow a well-known
 two-well potential nucleation problem
arises from the master equation, which analysis is made based on general well-known  methods of statistical physics~\cite{Gardiner,vanKampen}.
One of the main results of this analysis is the nucleation rate for the critical vehicle cluster (critical nuclei) whose occurence leads to a phase transition within the initial metastable state of traffic flow. As usual for other metastable systems of statistical physics~\cite{Gardiner,vanKampen},
in the metastable traffic flow the nucleation rate for a phase transition is an exponential function of a control parameter, 
flow rate (or vehicle density) for traffic flow~\cite{Mahnke1997,Mahnke1999,Kuehne2002,Kuehne2004,MahnkeRev}.
Rather than these common well-known results, an interest for the field 
has a nucleation model for a metastable traffic flow. Both the model and associated dependencies of the nucleation rates for phase transitions on
traffic variables and control parameters  should be adequate with   real traffic flow features. 

In~\cite{Mahnke1997,Mahnke1999,Kuehne2002},   models for vehicle cluster nucleation in  an initial spatially homogeneous traffic flow  on a homogeneous circular  road are introduced (see, however, Sect.~\ref{Comparison_S}). 
 One of the basic assumptions of these models is that in an initial homogeneous free flow firstly 
 random precluster should emerge from fluctuations. In other words,
 this precluster foregoes subsequent cluster evolution towards  a critical cluster (critical nuclei)
 whose growth leads to a phase transition. 
  The  rate of  precluster emergence  $w_{+}(0)$ in  traffic, in which initially no vehicle cluster exists,
  can be different from the attachment rate of cluster evolution, when a random cluster with $n\geq 1$ vehicles  already exists at the road~\cite{Mahnke1997,Mahnke1999,Kuehne2002,MahnkeRev}.

 A first nucleation model based on the master equation 
 for traffic breakdown at an on-ramp bottleneck, i.e., in an open traffic system  
 has been suggested by K\"uhne et al.~\cite{Kuehne2004} and Mahnke et al. (Chap.~17 of Ref.~\cite{MahnkeRev}). 
As in the models of homogeneous road~\cite{Mahnke1997,Mahnke1999,Kuehne2002}, 
in this model at given flow rates to the on-ramp $q_{\rm on}$ and on the main road upstream of the 
 on-ramp $q_{\rm in}$ random vehicle cluster occurrence and evolution that can lead to traffic breakdown
 are realized    {\it only}    after a vehicle precluster consisting of one vehicle randomly appears at the bottleneck. 
 The   rate of   precluster emergence is equal to the flow rate to the on-ramp~\cite{Kuehne2004,MahnkeRev}:
 $w_{+}(0)=q_{\rm on}$. This idea about foregoing precluster emergence
    is associated with a basic model assumption
 that at $q_{\rm on}=0$ no vehicle cluster can randomly appear, specifically no traffic breakdown is possible~\cite{Kuehne2004,MahnkeRev}.
  Probably, this basic model assumption leads to  the  nucleation rate for traffic breakdown that
  is proportional to   $q_{\rm on}$~\cite{Kuehne2004,MahnkeRev}.
  However, in real traffic flow both
  flow rates $q_{\rm on}$ {\it and} $q_{\rm in}$ exhibit random fluctuations that can cause cluster emergence.

    Moreover, whereas the basic assumption
about the necessity of precluster emergence 
is physically justified for a homogeneous road~\cite{Mahnke1997,Mahnke1999,Kuehne2002}, this is not the case for 
 for traffic breakdown at  the bottleneck.
This is because
in contrast with the model of a homogeneous road~\cite{Mahnke1997,Mahnke1999,Kuehne2002}, an initial state
of free flow at the bottleneck at $q_{\rm on}>0$ and $q_{\rm in}>0$  is {\it non-homogeneous
regardless of fluctuations}~\cite{KernerBook}. This means that even there were no fluctuations  in free flow at the bottleneck, nevertheless
free   flow is non-homogeneous in a neighborhood of the bottleneck.
This is because two different flows permanent merge within the merging region of the on-ramp -- the flow with the
rate $q_{\rm on}$ and the flow with the rate $q_{\rm in}$.

It has been shown in three-phase traffic theory~\cite{KernerBook} that due to the merging of these flows a steady state
of free flow at the bottleneck is non-homogeneous: A permanent and motionless local perturbation (deterministic perturbation) occurs at
the bottleneck. Within this perturbation that can be considered a deterministic vehicle cluster the speed is lower
and density is greater than downstream of the cluster.
    This deterministic cluster can lead to traffic breakdown even if there were {\it no}
  random fluctuations at the bottleneck~\cite{Kerner2000A,KernerBook}. 
  However, in the nucleation model~\cite{Kuehne2004,MahnkeRev} the master equation is formulated for probability of
  a vehicle cluster, which can occur due to fluctuations {\it only}. We see that for an initially non-homogeneous
  traffic flow, which occurs at the bottleneck, another physical approach should be applied. 
  This is due to at least two reasons mentioned above: (i)
  Regardless of fluctuations, there is already a vehicle cluster at the bottleneck that growth can lead to traffic breakdown: No vehicle
  precluster emerging from fluctuations is necessary for traffic breakdown. (ii) When influence of fluctuations on traffic breakdown is considered,
  fluctuations caused by both flow rates $q_{\rm on}$  and $q_{\rm in}$ should be taken into account. 
    
  Thus, in contrast with the nucleation model at the bottleneck of Ref.~\cite{Kuehne2004,MahnkeRev}
  a nucleation model that is adequate with empirical results for traffic breakdown
  at the bottleneck should consider vehicle cluster evolution in an initially non-homogeneous free traffic. 
  This random vehicle cluster should include {\it both}
  the total vehicle number   within an initial deterministic cluster {\it and} addition vehicles within the cluster, which occur
  due to fluctuations.
  
  However, there are no nucleation models and associated probabilistic theories for an initially 
  spatially non-homogeneous free  flow.
  Moreover, structure non-homogeneities play also
  a very important role for phase transitions
   in many other multiple-particle systems of statistical physics like
  granular flow, biological physics, reaction-diffusion systems, etc. (see references, e.g. in~\cite{SW1,SW2,SW3,SW4,SW5,KO,Michailov1}).
  Thus, it seems
  important  to derive a probabilistic theory for traffic breakdown in non-homogeneous traffic flow. 
 
  In this paper, a  nucleation model for the probabilistic breakdown phenomenon in an initially 
  spatially non-homogeneous traffic  flow at an on-ramp bottleneck is presented. 
The article is organized as follows.
In Sect.~\ref{Sect_prob}, the nucleation model is   considered. Nucleation rate and the mean time delay
for traffic breakdown that result from this model 
are studied in Sect.~\ref{Sect_time}.
  Results of the paper 
  as well as their   comparison with earlier nucleation models for traffic breakdown
  of Ref.~\cite{Mahnke1997,Mahnke1999,Kuehne2002,Kuehne2004,MahnkeRev}  are discussed in Sect.~\ref{Sect_Dis}.

\section{Nucleation model of  traffic breakdown in initially non-homogeneous free flow at on-ramp bottleneck}
\label{Sect_prob}

\subsection{Non-homogeniety in free flow at bottleneck as reason for vehicle cluster \label{Nonhom_S}}

\begin{figure}
\begin{center}
\includegraphics[width=7 cm]{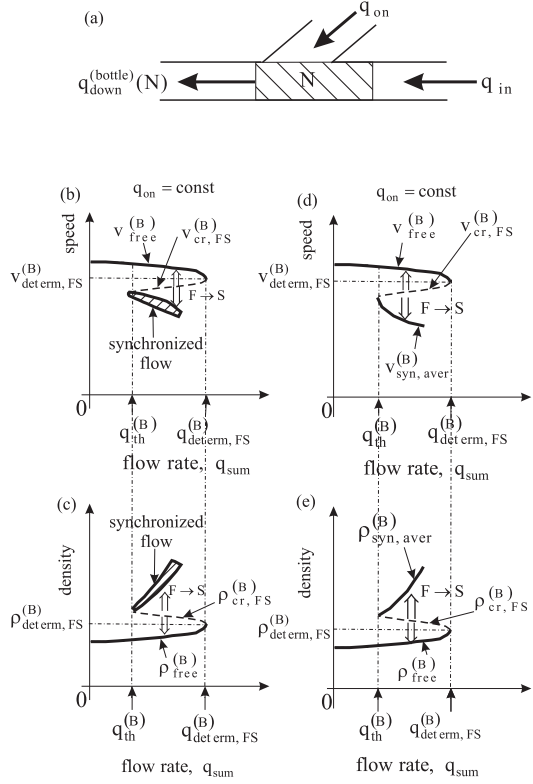}
\caption{Explanation of the basis of nucleation model: (a) Sketch of an on-ramp bottleneck.
(b, c) --
Z-shaped speed--flow (a) and the associated S-shaped density--flow characteristics for an F$\rightarrow$S 
  transition. (d, e) -- Simplified  Z-shaped speed--flow (d) and   S-shaped density--flow characteristics (e)
related to (b) and (c), respectively.
\label{FS_Intr} } 
\end{center}
\end{figure}

In accordance with three-phase traffic theory~\cite{Kerner2000A,KernerBook}, in a nucleation model 
we assume that
the breakdown phenomenon at an on-ramp bottleneck is associated
 with an initial non-homogeneity of  free flow at the bottleneck. This non-homogeneity exists at the bottleneck
 even in a hypothetical case in which there were no fluctuations in traffic flow. In this hypothetical case, this non-homogeneity
 can be considered
 a  {\it deterministic} (permanent) local perturbation localized at the bottleneck.
 The non-homogeneity of an initial free flow at the bottleneck is caused by two
 flows, which merge at the bottleneck: (i) An on-ramp inflow with the rate $q_{\rm on}$. (ii) A flow
 on the main road upstream of the bottleneck with the rate $q_{\rm in}$. This flow merging occurs
 permanent and on the same freeway location (within an on-ramp merging region). For this reason, the 
 non-homogeneity of  free flow at the bottleneck is motionless and permanent (Fig.~\ref{FS_Intr} (a)).
 
To explain features of this non-homogeneity, note that at a given high enough flow rate $q_{\rm in}$ in free flow on the main road upstream
 of the bottleneck, vehicles that merge from the on-ramp onto the main road force the vehicles on the main road to decelerate in the
 vicinity of an on-ramp merging region. This deceleration leads to a local decrease in speed 
 and consequently to a local increase in density 
 in the vicinity of the bottleneck. As a result, a local perturbation, i.e., non-homogeneity in free flow appears. 
 
 If no fluctuations in free flow are considered, as mentioned above the non-homogeneity in free flow
 in the form of a deterministic local perturbation occurs at the bottleneck.
 The speed  $v^{\rm (B)}_{\rm free}$
 and density $\rho^{\rm (B)}_{\rm free}$ within this deterministic perturbation correspond to the conditions 
  \begin{equation}
v^{\rm (B)}_{\rm free}<v^{\rm (free)}, \ \rho^{\rm (B)}_{\rm free}>\rho^{\rm (free)},
 \label{v_free_rho_free}
 \end{equation}
 where $v^{\rm (free)}$ and  $\rho^{\rm (free)}$ are the average vehicle speed and density in homogeneous free flow
 on the main road downstream of the  perturbation (Fig.~\ref{FS_Intr} (a)).
 Because  the  deterministic
 local perturbation is  permanent and  motionless,
 at given $q_{\rm in}$ and $q_{\rm on}$, the total flow rate 
 does not depend on the spatial co-ordinate, i.e., the speed  and density   on the main road  
 satisfy the following condition:
 \begin{equation}
q_{\rm sum}=v^{\rm (free)}\rho^{\rm (free)}=v^{\rm (B)}_{\rm free}\rho^{\rm (B)}_{\rm free},
 \label{q_sum_x}
 \end{equation}
  where
     \begin{eqnarray}
q_{\rm sum}=q_{\rm in}+q_{\rm on}.
 \label{inflow_sum}
 \end{eqnarray}
 The  inhomogeneous steady state
 within the deterministic local perturbation  
 can be considered   $\lq\lq$deterministic vehicle cluster" in free flow localized at the bottleneck or $\lq\lq$deterministic   cluster"
 for short.

When   $q_{\rm on}$ is a given value and the total flow rate $q_{\rm sum}$ increases, then   the vehicle speed $v^{\rm (B)}_{\rm free}$ within the deterministic
 cluster decreases and in accordance with
(\ref{q_sum_x}) the associated density $\rho^{\rm (B)}_{\rm free}$ increases. However, this increase is limited by some critical
 density $\rho^{\rm (B)}_{\rm free}=\rho^{\rm (B)}_{\rm determ, \ FS}$  (Fig.~\ref{FS_Intr} (c))
 within the deterministic cluster
 associated with a critical flow rate   
  \begin{equation}
 q_{\rm sum}=q^{\rm (B)}_{\rm determ, \ FS}.
 \label{critical_determ} 
 \end{equation} 
 After this critical deterministic perturbation is reached,
 the further increase in $q_{\rm sum}$  leads to {\it  deterministic traffic breakdown}   at the bottleneck causing  spontaneous synchronized flow
 emergence at the bottleneck.  The critical deterministic perturbation
 can be considered  $\lq\lq$critical deterministic vehicle cluster"   ($\lq\lq$critical deterministic  cluster" for short)
  or  $\lq\lq$deterministic  nuclei for traffic breakdown". After the critical deterministic cluster  is reached,
  deterministic traffic breakdown occurs at the bottleneck even if there were no random perturbations in traffic flow at the bottleneck.

 Random perturbations within the initial deterministic cluster can cause  random traffic breakdown
 (F$\rightarrow$S transition) at the flow rate 
  \begin{equation}
q_{\rm sum}<q_{\rm determ, \ FS}^{\rm (B)},
 \label{random_FS}
 \end{equation}
 i.e., before the deterministic  nuclei for traffic breakdown  is reached. In this case,
 random  traffic breakdown nucleation can occur at the bottleneck (arrows labeled F$\rightarrow$S in Fig.~\ref{FS_Intr} (b, c)).  
 This is realized,  if through a   random increase in vehicle number
   within  the initial deterministic   cluster, the  amplitude of the resulting $\lq\lq$random vehicle cluster"
   ($\lq\lq$random   cluster" for short)
 exceeds some critical amplitude associated with
 a critical density within the random cluster $\rho^{\rm (B)}_{\rm cr, \ FS}$  
 (Fig.~\ref{FS_Intr} (c)).
 The  random cluster with
 the critical density 
  $\rho^{\rm (B)}_{\rm cr, \ FS}$ can be considered $\lq\lq$critical random vehicle cluster"
 at the bottleneck ($\lq\lq$critical random   cluster" for short) or $\lq\lq$random nuclei for traffic breakdown".
 If the amplitude of a random cluster is smaller than the critical one, the random  cluster decays towards the initial deterministic cluster.
 
 \subsection{Master equation}
 
 \subsubsection{Basic assumptions for master equation \label{Assum_Sect}}

There are the following basic assumptions for the nucleation model in an initial non-homogeneous free flow at the bottleneck:

(i) There is no vehicle precluster, which random occurrence through a random decrease in speed of one of the vehicles
in the initial free flow should be necessary for vehicle cluster emergence and
subsequent   cluster evolution. Traffic breakdown occurs within a deterministic  vehicle cluster that is motionless and permanently exists
in a neighborhood of the bottleneck. 

\begin{figure}
\begin{center}
\includegraphics[width=10 cm]{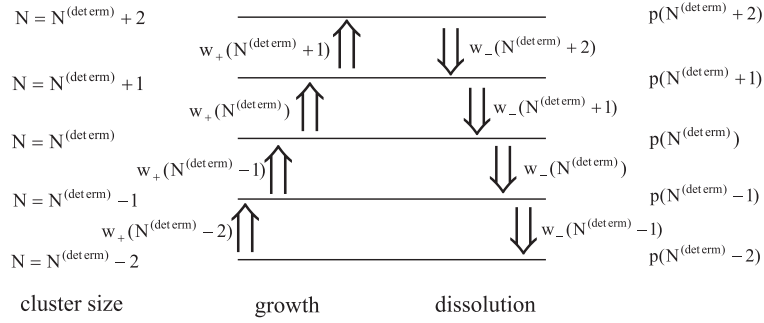}
\caption{Schematic illustration of cluster transformation in an initially non-homogeneous free flow at the bottlneck.
\label{Cluster} } 
\end{center}
\end{figure}

(ii) A master equation should describe probability $p$ for random vehicle cluster evolution in which the cluster size $N$, 
i.e., the total vehicle number within the motionless vehicle cluster
 randomly changes due to fluctuations in a neighborhood of the deterministic cluster (Fig.~\ref{Cluster}).
 The size of the deterministic cluster $N^{\rm (determ)}$ does not depend on fluctuations in traffic flow.

(iii) The attachment rate $w_{+}$ onto this vehicle cluster is not a function of the cluster size $N$, i.e. 
$w_{+}(0)=w_{+}(N)=q_{\rm in}+q_{\rm on}$.

(iv) A dependence of the detachment rate $w_{-}$ from a cluster is a non-linear function on $N$, which consists of at least two
branches associated with the deterministic cluster and a critical vehicle cluster (nuclei for traffic breakdown). 

 (v) The on-ramp inflow and the flow upstream of the bottleneck make a different influence on the cluster size. To take into account
 this assymetric behavior in the model, the detachment rate $w_{-}$  also depends on
 $q_{\rm on}$.

The  assumption (i) can be explained by existence of  a deterministic vehicle cluster at the bottleneck. This cluster can lead to traffic breakdown 
 at the bottleneck even there were no fluctuations in traffic flow, i.e., no precluster is necessary for vehicle cluster occurrence in
 an initially steady state of non-homogeneous free flow at the bottleneck.
 To explain the assumption (ii), 
note that real random fluctuations can cause either a decrease or an increase in the cluster size $N$ and, respectively,
either a decrease or an increase in  vehicle density within the cluster. 
Real  random fluctuations, which decrease the cluster size $N$, are associated with an increase in  speed within the cluster (Sect.~\ref{Nonhom_S}). Therefore, these fluctuations
 can prevent the deterministic breakdown. In contrast, random fluctuations, which increase the cluster size $N$, i.e., 
 decrease the speed within the cluster,
 can cause traffic breakdown  
before the deterministic  nuclei for traffic breakdown  is reached, i.e., when the condition  (\ref{random_FS})
is satisfied. Regardless of the cluster size $N$, the attachment rate into the cluster is determined by fluctuations in both flow rates
$q_{\rm in}$ {\it and} $q_{\rm on}$. This explains the assumption (iii).
 Grounds for the assumptions (iv) and (v) appear in Sect.~\ref{N_grounds}. 

  We consider the dynamics of  the total vehicle number $N$ within a vehicle cluster localized at the bottleneck
   (dashed region in Fig.~\ref{FS_Intr} (a)). 
  It is assumed that within the cluster free flow is spatially non-homogeneous, specifically the 
   region of spatial cluster localization is bounded upstream and downstream with homogeneous free flows.
The total number of vehicles $N$ 
  within  the cluster can either increase or decrease over time randomly  
  in comparison with a value $N=N^{\rm (determ)}$ for the case in which
  the deterministic cluster exists at the bottleneck only.

 In accordance with the above basic assumptions, 
 the probability $p(N,t)$  to find $N$ vehicles within the cluster at the bottleneck  reads as follows 
\begin{eqnarray}
\label{Prob}
\frac{\partial  p(N,t)}{\partial t}=  w_{+}(N-1) p(N-1,t) +w_{-}(N+1)p(N+1,t) \nonumber
 \\
 -[w_{+}(N) +w_{-}(N)] p(N,t),  \quad  \textrm{at $N > 0$},  
\end{eqnarray} 
\begin{eqnarray}
\frac{\partial p(0,t)}{\partial t}=w_{-}(1) p(1,t) -w_{+}(0)p(0,t), \quad  \textrm{at $N = 0$}, \quad
\label{Prob0}
\end{eqnarray}
with the boundary condition
\begin{eqnarray}
w_{-}(0)=0,
\label{Prob0_boundary}
\end{eqnarray}
 In accordance with the assumption (3), 
 the vehicle attachment rate  $w_{+}$  is  independent of  $N$, i.e.,
\begin{eqnarray}
w_{+}=q_{\rm sum}.
 \label{inflow}
 \end{eqnarray}

 \subsubsection{Vehicle detachment rate from cluster \label{N_grounds}}
 
The vehicle detachment rate    $w_{-}$ is obviously
equal to the outflow rate from the cluster
  \begin{eqnarray}
w_{-}(N)=q^{\rm (bottle)}_{\rm down}(N).
 \label{w_}
 \end{eqnarray}
 To understant a qualitative shape of the function $q^{\rm (bottle)}_{\rm down}(N)$, note that in accordance with three-phase traffic theory~\cite{KernerBook},
 speed states within the deterministic   cluster $v^{\rm (B)}_{\rm free}(q_{\rm sum})$,
 the speed within the critical random   cluster $v^{\rm (B)}_{\rm cr, \ FS}(q_{\rm sum})$,  along with
 a 2D  
 synchronized flow speed states~\cite{KernerBook}, together form a Z-shaped speed--flow characteristic for  traffic breakdown
 (Fig.~\ref{FS_Intr} (b)).
 The associated density--flow characteristic, which consists of density states
 within the deterministic   cluster $\rho^{\rm (B)}_{\rm free}(q_{\rm sum})$,
 the density within the critical random cluster $\rho^{\rm (B)}_{\rm cr, \ FS}(q_{\rm sum})$,  along with
 a 2D  
 synchronized flow  states, has obviously a S-shaped form 
 (Fig.~\ref{FS_Intr} (c))~\footnote{Within the deterministic cluster the speed $v^{\rm (B)}_{\rm free}$ and density
 $\rho^{\rm (B)}_{\rm free}$ shown in 
 Fig.~\ref{FS_Intr} (b--e) are connected by the formula  (\ref{q_sum_x}). The number of vehicles within the
 cluster $N=\int{\rho dx}$, where integration is performed over the region of   cluster localization.} 
Due to an F$\rightarrow$S transition,  fast cluster growth   
at the bottleneck occurs leading to congested pattern formation,
 i.e., either a synchronized flow  pattern (SP) or a general pattern (GP) appears upstream of the bottleneck~\cite{KernerBook}.
 However, the nucleation effect leading to traffic breakdown and its characteristics are fully independent of 
  possible congested patterns  resulting
from this F$\rightarrow$S transition. For this reason,
we can average the infinity of   synchronized flow states (dashed regions in (b, c)) 
for each given flow rate $q_{\rm sum}$
   to one speed $v^{\rm (B)}_{\rm syn, \ aver}(q_{\rm sum})$ and to one density
   $\rho^{\rm (B)}_{\rm syn, \ aver}(q_{\rm sum})$ (Fig.~\ref{FS_Intr} (d, e))~\footnote{Concerning
  synchronized flow states in the vicinity of   the bottleneck associated with a resulting congested pattern,   we should note that 
assumptions
     used in our nucleation model  are satisfied only for a localized SP (LSP)~\cite{KernerBook}.
     An LSP is an SP whose downstream front   is fixed at the bottleneck. The upstream front of the LSP is localized
     on the main road  at 
     some finite distance upstream of the bottleneck, i.e., the width (in the longitudinal direction) of the LSP is always limited.
     Synchronized flow states   at the bottleneck are drawn in Fig.~\ref{FS_Intr} (b, c)
and  in other illustrations  only for the case in which
  the congested pattern   is an LSP.}. A consideration of the resulting congested patterns  
  is beyond the scope of this article.
 
\begin{figure}
\begin{center}
\includegraphics[width=7 cm]{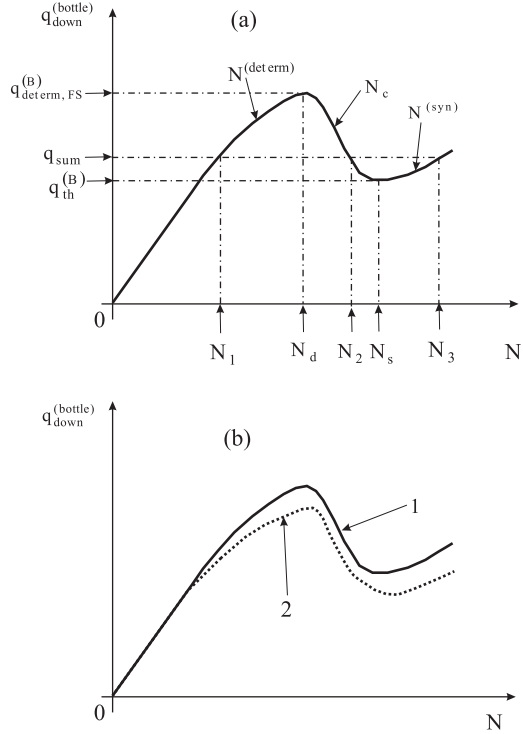}
\caption{Qualitative dependencies of the outflow rate $q^{\rm (bottle)}_{\rm down}$
on the total vehicle number $N$ within the cluster localized  at the bottleneck (a), and possible 
dependencies of the N-shaped function $q^{\rm (bottle)}_{\rm down}(N)$ on $q_{\rm on}$
 for two different values $q_{\rm on}$ (b); curve 1 for $q_{\rm on}=q^{(1)}_{\rm on}$, curve 2 for $q_{\rm on}=q^{(2)}_{\rm on}>q^{(1)}_{\rm on}$.
\label{q_rho} } 
\end{center}
\end{figure}

In the model, it is assumed that the
 shape of the characteristic $q^{\rm (bottle)}_{\rm down}(N)$ (Fig.~\ref{q_rho}) follows from the S-shaped density--flow characteristic 
 of three-phase traffic theory (Fig.~\ref{FS_Intr} (e)): The characteristic $q^{\rm (bottle)}_{\rm down}(N)$
  has at least two different branches $q^{\rm (bottle)}_{\rm down}(N)$  labeled $N^{\rm (determ)}$ and
 $N_{\rm c}$ in Fig.~\ref{q_rho} (a). These branches  are related to the vehicle number ranges, respectively,
 given by the conditions 
 \begin{eqnarray}
0\leq N \leq N_{\rm d}
 \label{range1}
 \end{eqnarray}
 and
  \begin{eqnarray}
N_{\rm d}< N \leq N_{\rm s}.
 \label{range2}
 \end{eqnarray} 
 The branches $N^{\rm (determ)}$ and
 $N_{\rm c}$ in Fig.~\ref{q_rho} (a) are associated with the density branches
 $\rho^{\rm (B)}_{\rm free}$ and $\rho^{\rm (B)}_{\rm cr, \ FS}$
 of the S-shaped density--flow characteristic in Fig.~\ref{FS_Intr} (e), respectively. The branch
 $N^{\rm (determ)}$ is associated with the case in which at   a high
 enough flow rate $q_{\rm sum}$  and the on-ramp flow rate  $q_{\rm on}>0$ the deterministic cluster exists at the bottleneck.
 The branch
 $N_{\rm c}$ is associated with the case in which the critical random cluster whose growth leads to an F$\rightarrow$S transition
 occurs at the bottleneck.
 
In addition, from the S-shaped density--flow characteristic 
 (Fig.~\ref{FS_Intr} (e)) can be seen that for the case in which an LSP results from an F$\rightarrow$S transition,
at
  \begin{eqnarray}
N>N_{\rm s},
 \label{range3}
 \end{eqnarray}
 there is a third branch $N^{\rm (syn)}$ on the characteristic $q^{\rm (bottle)}_{\rm down}(N)$ (Fig.~\ref{q_rho} (a))
 associated with the branch
  $\rho^{\rm (B)}_{\rm syn, \ aver}$ for averaged synchronized flow states in Fig.~\ref{FS_Intr} (e).
  In this case,   $q^{\rm (bottle)}_{\rm down}(N)$  (\ref{w_}) is a N-shaped flow--vehicle-number characteristic.

 At the critical point $N=N_{\rm d}$ at which the branches  $N^{\rm (determ)}$
 and  $N_{\rm c}$ merges, the function $q^{\rm (bottle)}_{\rm down}(N)$ has its maximum point. 
 At the threshold point $N=N_{\rm s}$ at which the branches  $N_{\rm c}$
 and  $N^{\rm (syn)}$ merges, the function $q^{\rm (bottle)}_{\rm down}(N)$ has its minimum point.

Quantitative characteristics of the N-shaped function $q^{\rm (bottle)}_{\rm down}(N)$ (e.g., values $N_{\rm d}$
and $N_{\rm s}$)
can depend on the flow rate to the on-ramp
 $q_{\rm on}$ (assumption (v) in Sect.~\ref{N_grounds}). This is because  the on-ramp inflow
 and the flow upstream of the bottleneck can make a considerable different influence
 on the cluster size and  the outflow rate from the cluster $q^{\rm (bottle)}_{\rm down}$. In particlular, 
 it can   turn out that at the same $N$ the greater $q_{\rm on}$, the more difficult for vehicles to escape from
 the   cluster, i.e., the less  $q^{\rm (bottle)}_{\rm down}$ is.    This  is confirmed
 by microscopic simulations~\cite{KKW} and reflected in Fig.~\ref{q_rho} (b)
 in which it is assumed that the greater $q_{\rm on}$, the greater $N_{\rm d}$
and $N_{\rm s}$ are. Thus, in a general case instead of
 (\ref{w_}) we should use 
 \begin{equation}
w_{-}(N)=q^{\rm (bottle)}_{\rm down}(N, \ q_{\rm on}).
\label{w_final}
 \end{equation}
 A possible impact of the flow rate  $q_{\rm on}$
 on quantitative characteristics of the mean time delay for an F$\rightarrow$S transition is discussed in Sect.~\ref{Sect_Dis}.
 
 Through the use of the basic assumptions (i)-(v) of Sect.~\ref{Assum_Sect} and the chosen shape of the function (\ref{w_})
 discussed above (Fig.~\ref{q_rho} (a)), critical cluster occurrence describes an F$\rightarrow$S transition at the bottleneck.
 There are two reasons for this statement: (i) A vehicle cluster is motionless, i.e.,  fixed at the bottleneck. This is related to the 
 definition of synchronized flow whose downstream front is usually fixed at the bottleneck, whereas the downstream from of a wide moving jam
 propagates trough any bottleneck while maintaining the mean downstream jam velocity~\cite{KernerBook}. (ii) The shape of the chosen function (\ref{w_})
 in the nucleation model 
 associated with this motionless cluster (Fig.~\ref{q_rho} (a)) follows from a Z-shaped speed--flow characteristic for  traffic  breakdown
 (Fig.~\ref{FS_Intr} (b)) for an F$\rightarrow$S transition at the bottleneck
 found in a microscopic traffic flow theory. In this theory has been shown that if these two requirements are satisfied, then rather than
 an F$\rightarrow$J transition (moving jam emergence)  an F$\rightarrow$S transition, i.e., synchronized flow emergence occurs at the bottleneck. As follows from this microscopic theory, after the F$\rightarrow$S transition has already occurred, moving jams
 can emerge in this synchronized flow. However, the nucleation model describes traffic breakdown,
 i.e., an F$\rightarrow$S transition, specifically the rate of traffic breakdown (synchronized flow) nucleation {\it only}. 

In addition, it should be noted that the branch for synchronized flow $v^{\rm (B)}_{\rm syn, \ aver}$ in Figs.~\ref{FS_Intr} (b-e)
and the associated branch 
 $N^{\rm (syn)}$ on the characteristic $q^{\rm (bottle)}_{\rm down}(N)$ (Fig.~\ref{q_rho} (a)) follow from the microscopic traffic theory, 
 rather than from the nucleation model. This branch, which is shown only with the aim of a qualitative illustration of a possible traffic flow state after synchronized flow nucleation,  has no influence  on the nucleation rate of an F$\rightarrow$S transition.

  \subsection{Steady states}
 
Steady states of vehicle number $N$ at given $q_{\rm in}$ and $q_{\rm on}$  
are associated with solutions of the equation
 \begin{eqnarray}
w_{+}=w_{-}(N).
 \label{steady_states}
 \end{eqnarray}
In accordance with  (\ref{inflow}), (\ref{w_final}), they
are found from the condition
\begin{eqnarray}
q_{\rm sum}=q^{\rm (bottle)}_{\rm down}(N, \ q_{\rm on}).
 \label{steady_states_q}
 \end{eqnarray}
 As can be seen from Fig.~\ref{q_rho} (a),
at given flow rates $q_{\rm on}$ and $q_{\rm in}$ that satisfy the condition
\begin{equation}
q^{\rm (B)}_{\rm th}<q_{\rm sum}<q^{\rm (B)}_{\rm determ, \ FS}
\label{q_sum_cond}
\end{equation}
there can be at least two steady states:
$N=N_{1}$ associated with 
 the deterministic cluster and  $N=N_{2}$ associated with
 the critical random cluster.   These steady states are  the roots of Eq.  (\ref{steady_states_q}), i.e., they are associated with the
intersection points of the
horizontal line $q^{\rm (bottle)}_{\rm down}=q_{\rm sum}$ with the
 branches $N^{\rm (determ)}$ and
 $N_{\rm c}$ of the  characteristic 
$q^{\rm (bottle)}_{\rm down}(N, \ q_{\rm on})$
 (Fig.~\ref{q_rho} (a)), respectively. In addition, if an LSP occurs as a result of an F$\rightarrow$S transition,
 then there is a third root of Eq. (\ref{steady_states_q}), $N=N_{3}$, associated with
 the
intersection point of the
horizontal line $q^{\rm (bottle)}_{\rm down}=q_{\rm sum}$ with the
 branch  $N^{\rm (syn)}$ of the  characteristic $q^{\rm (bottle)}_{\rm down}(N)$.
 
 If the flow rate $q_{\rm sum}$ increases, then the   critical vehicle number difference within the cluster  
 \begin{equation}
 \Delta N_{\rm c}=N_{2}-N_{1} 
 \label{N_2_N_1}
 \end{equation}
 decreases. This   critical vehicle number difference is associated with the vehicle number difference within the critical random cluster
   and within the initial
  deterministic cluster at the bottleneck. The growth of the critical random cluster leads to traffic breakdown at the bottleneck.
 
 At the critical flow rate (\ref{critical_determ}),
we get $\Delta N_{\rm c}=0$:  The steady states $N_{1}$ and $N_{2}$
 merge into one point with the critical vehicle number $N=N_{\rm d}$ at which 
 $q^{\rm (B)}_{\rm determ, \ FS}=q^{\rm (bottle)}_{\rm down}(N_{\rm d}, \ q_{\rm on})$. 
 At $q_{\rm sum}\geq q^{\rm (B)}_{\rm determ, \ FS}$
  the deterministic traffic breakdown should occur
 even if there is no random increase in the vehicle number within the initial deterministic cluster at the bottleneck. 
  
If the flow rate $q_{\rm sum}$ decreases gradually, then the   threshold flow rate 
   \begin{equation}
 q_{\rm sum}=q^{\rm (B)}_{\rm th} 
 \label{critical_thresh}
 \end{equation}
is reached at which the steady states $N_{2}$ and $N_{3}$
 merge into one threshold steady state   $N=N_{\rm s}$ at which 
 $q^{\rm (B)}_{\rm th}=q^{\rm (bottle)}_{\rm down}(N_{\rm s}, \ q_{\rm on})$.
 
 \section{Nucleation rate of traffic breakdown at bottleneck} 
 \label{Sect_time}
 
  As follows from the analysis of the model (\ref{Prob})-(\ref{inflow}), (\ref{w_final}) (Appendix~\ref{App}), in the   flow rate range
  (\ref{q_sum_cond})
 the mean time delay of an F$\rightarrow $S transition  at the bottleneck is
\begin{eqnarray}
T^{\rm (B, \ mean)}_{\rm FS}=C \exp{\big \{\Delta \Phi \big \}}, 
\label{time_FS}
\end{eqnarray}
where a potential
 barrier
\begin{eqnarray}
\Delta \Phi= \Phi(N_{2})- \Phi(N_{1}),
\label{delta_pi}
\end{eqnarray}
the   potential $\Phi(N)$ is
\begin{eqnarray}
\label{potential_eq}
\Phi(N)=\left\{
\begin{array}{ll}
\sum^{N}_{n=1} \ln{\frac{w_{-}(n)}{w_{+}}}  &  \textrm{at $N >0$}, \\
0 &  \textrm{at $N=0,$}
\end{array} \right.
\end{eqnarray}
\begin{eqnarray}
C=2\pi \Big(w^{\prime}_{-}(N_{1}) 
\mid w^{\prime}_{-}(N_{2})\mid \Big)^{-\frac{1}{2}}, 
\label{factor}
\end{eqnarray}
  $w^{\prime}_{-}(N)=dw_{-}/dN$.
  Respectively, the   nucleation rate for traffic breakdown 
at the bottleneck is
 \begin{equation}
G^{\rm (B)}_{\rm FS}=\frac{1}{T^{\rm (B, \ mean)}_{\rm FS}}=C^{-1} \exp{\big \{-\Delta \Phi \big \}}.
 \label{Probab_FS}
 \end{equation}

 \begin{figure}
\begin{center}
\includegraphics[width=7 cm]{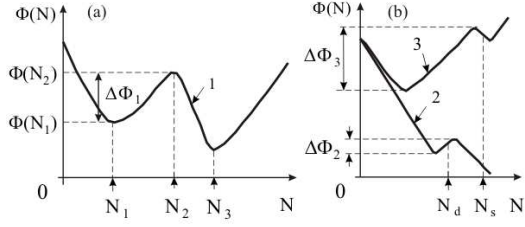}
\caption{Qualitative shape of the   potential  $\Phi(N)$ (\ref{potential_eq}) for different
 flow rates $q_{\rm sum}$: curves 1, 2, and 3 are related to the corresponding flow rates
   $q^{(1)}_{\rm sum}$, $q^{(2)}_{\rm sum}$, and $q^{(3)}_{\rm sum}$ satisfying the condition
   $q^{(3)}_{\rm sum}<q^{(1)}_{\rm sum}<q^{(2)}_{\rm sum}$.
\label{potential} } 
\end{center}
\end{figure}

To find a qualitative shape  $\Phi(N)$ (\ref{potential_eq}) (Fig.~\ref{potential}), a change 
 in  $\Phi(N)$ 
between two neighboring points $N$ and $N-1$ that equals  
\begin{eqnarray}
\label{potential_change}
\delta \Phi(N)=\Phi(N)-\Phi(N-1)= {\ln \frac{w_{-}(N)}{w_{+}}}
\end{eqnarray}
can be used.
The  value $\delta \Phi(N)$ (\ref{potential_change}) becomes zero at the maximum and minimum points of the function $\Phi(N)$, i.e., at
 the roots of Eq.  (\ref{steady_states}) that are the points $N=N_{\rm i}, \ i=1,2,3$ discussed above (Fig.~\ref{q_rho} (a)).
 The value $\delta \Phi(N)>0$ at $w_{-}(N)>w_{+}$, i.e., at points of the curve $w_{-}(N)$
above the horizontal line $q=q_{\rm sum}$ in (Fig.~\ref{q_rho} (a)). In contrast,
$\delta \Phi(N)<0$ at $w_{-}(N)<w_{+}$, i.e., at points of the curve $w_{-}(N)$
below the horizontal line $q=q_{\rm sum}$.
 
 It can be seen from (\ref{time_FS})   that the mean time delay for traffic breakdown decreases exponentionally
 with increase in  potential
 barrier $\Delta \Phi$
(\ref{delta_pi}).    If in
Fig.~\ref{potential} the total flow rate increases from
$q_{\rm sum}=q^{(1)}_{\rm sum}$ to $q_{\rm sum}=q^{(2)}_{\rm sum}$, which is close to the
critical flow rate  (\ref{critical_determ}) for  deterministic traffic breakdown, then  the potential
 barrier $\Delta \Phi$
(\ref{delta_pi}) decreases from $\Delta \Phi_{1}$ to $\Delta \Phi_{2}$. 

In contrast, if the total flow rate decreases from
$q_{\rm sum}=q^{(1)}_{\rm sum}$ to $q_{\rm sum}=q^{(3)}_{\rm sum}$, which is close to the
threshold flow rate $q^{(B)}_{\rm th}$ (\ref{critical_thresh}) for random traffic breakdown, then  the potential
 barrier $\Delta \Phi$
(\ref{delta_pi}) increases from $\Delta \Phi_{1}$ to $\Delta \Phi_{3}$ (Fig.~\ref{potential}).  
At
the threshold point $q_{\rm sum}=q^{\rm (B)}_{\rm th}$ (\ref{critical_thresh}),
 the potential
 barrier  $\Delta \Phi(N)$ 
reaches the maximum value
\begin{eqnarray}
\Delta \Phi= \Phi(N_{s})- \Phi(N_{\rm th}),
\label{potential_th}
\end{eqnarray}
where  $N_{\rm th}=N_{1}$ at $ q_{\rm sum}=q^{\rm (B)}_{\rm th}$.
 As a result, 
the mean time delay $T^{\rm (B, \ mean)}_{\rm FS}$ (\ref{time_FS}) 
strongly increases  as $q_{\rm sum}$ approaches the threshold point $q^{\rm (B)}_{\rm th}$. 
Under the condition  
\begin{equation}
q_{\rm sum}<q^{\rm (B)}_{\rm th}
\label{no_speed_bre}
\end{equation}
 no traffic breakdown at the bottleneck regardless of a random increase in the vehicle number within the
 cluster is possible at the bottleneck.

If  in the vicinity of the critical vehicle number $N_{\rm d}$   the function $w_{-}(N)$ (\ref{w_final}) 
 can be approximated by a parabolic function of $N$, 
then the following approximate formula can be derived from (\ref{time_FS})  (Appendix~\ref{App}):
\begin{equation}
 T^{\rm (B, \ mean)}_{\rm FS} = 
\frac{\sqrt{2} \pi N_{\rm d}}{q^{\rm (B)}_{\rm determ, \ FS}(\xi_{\rm d} \Delta_{\rm c})^{1/2} }
\bigg(\frac{1+\Delta_{\rm c}^{1/2}}{1-\Delta_{\rm c}^{1/2}}\bigg) ^{2\sqrt{2/\xi_{\rm d}}N_{\rm d}}\exp{\bigg(-\frac{4\sqrt{2} N_{\rm d}{\Delta_{\rm c}^{1/2}}}{\sqrt{\xi_{\rm d}}} \bigg)},
\label{time_FS_approx_new}
\end{equation}
where 
\begin{equation}
\xi_{\rm d}=-(N^{2}d^{2}\ln{w_{-}}/dN^{2})\big| _{N=N_{\rm d}}
\label{xi_determ}
\end{equation}
is a dimensionless value of the order of 1, 
\begin{equation}
\Delta_{\rm c}=\frac{q^{\rm (B)}_{\rm determ, \ FS}- q_{\rm sum}}{q^{\rm (B)}_{\rm determ, \ FS}},
\label{overcr}
\end{equation}
i.e., $\Delta_{\rm c}$ is the relative difference between
the critical flow rate
$q^{\rm (B)}_{\rm determ, \ FS}$
for the deterministic F$\rightarrow$S transition and the total flow rate $q_{\rm sum}$ (\ref{inflow_sum}). 
If in (\ref{time_FS_approx_new}) $\Delta_{\rm c}\ll$ 1, then we get
\begin{equation}
 T^{\rm (B, \ mean)}_{\rm FS} = 
\frac{\sqrt{2} \pi N_{\rm d}}{q^{\rm (B)}_{\rm determ, \ FS}(\xi_{\rm d} \Delta_{\rm c})^{1/2} }
 \exp{\bigg(\frac{8 N_{\rm d}{\Delta_{\rm c}^{3/2}}}{3 \sqrt{2\xi_{\rm d}}} \bigg)}.
\label{time_FS_approx}
\end{equation}
Respectively, the   nucleation rate 
$G^{\rm (B)}_{\rm FS}=1/T^{\rm (B, \ mean)}_{\rm FS}$ 
for traffic breakdown 
at the bottleneck associated with (\ref{time_FS_approx}) is
 \begin{equation}
G^{\rm (B)}_{\rm FS} = \frac{ q^{\rm (B)}_{\rm determ, \ FS}(\xi_{\rm d} \Delta_{\rm c})^{1/2} }{\sqrt{2} \pi N_{\rm d}}
 \exp{\bigg(-\frac{8 N_{\rm d}{\Delta_{\rm c}^{3/2}}}{3 \sqrt{2\xi_{\rm d}}} \bigg)}.
 \label{probability_FS_approx}
 \end{equation}
Note that $q^{\rm (B)}_{\rm determ, \ FS}$, $N_{\rm d}$, and $\xi_{\rm d}$ depend $q_{\rm on}$. Therefore,
the mean time delay  $T^{\rm (B, \ mean)}_{\rm FS}$ (\ref{time_FS_approx})
and the nucleation rate $G^{\rm (B)}_{\rm FS}$ ( \ref{probability_FS_approx})
are functions of $q_{\rm sum}$ and $q_{\rm on}$.

If the flow rate $q_{\rm on}$ decreases continuously up to a small enough value
 (however, we assume that $q_{\rm on}>0$, i.e.,    the deterministic cluster
 still exists at the bottleneck), then the values $\xi_{\rm d}$, $N_{\rm d}$, 
and  $q^{\rm (B)}_{\rm determ, \ FS}$ in (\ref{overcr}), (\ref{probability_FS_approx}) and, therefore,
  the nucleation rate $G^{\rm (B)}_{\rm FS}$ ( \ref{probability_FS_approx})
do {\it not} decrease proportionally to this decrease in $q_{\rm on}$. In contrast, in this limit case
$\xi_{\rm d}\rightarrow \xi_{\rm d, \ lim}$, 
$N_{\rm d} \rightarrow N_{\rm d, \ lim}$,
and  $q^{\rm (B)}_{\rm determ, \ FS}\rightarrow q^{\rm (B)}_{\rm determ, \ lim}$, where $\xi_{\rm d, \ lim}$,
 $N_{\rm d, \ lim}$, and  $q^{\rm (B)}_{\rm determ, \ lim}$ are constants.
Taking into account that in this case in $\Delta_{\rm c}$ (\ref{overcr})
the flow rate $q^{\rm (B)}_{\rm determ, \ FS}\approx q^{\rm (B)}_{\rm determ, \ lim}={\rm const}$,
we can see that at   small enough  values of $q_{\rm on}$ the   nucleation rate for traffic breakdown (\ref{probability_FS_approx})
  depends  on  
the total flow rate $q_{\rm sum}$ only. In other words, in this limit case 
at a given $q_{\rm sum}$ within the flow rate range (\ref{q_sum_cond})
the   nucleation rate for   traffic breakdown at the bottleneck (\ref{probability_FS_approx})
 tends to a finite constant value,
 which is greater than zero (see Sect.~\ref{Num_Sym_S}).    
      
   When $q_{\rm on}= 0$,
   the road can be considered homogeneous one without bottlenecks. Then there 
   is no deterministic perturbation (deterministic cluster) at the bottleneck and, therefore, the nucleation model and results of this article  
   cannot be applied. 
     In three-phase traffic theory, the breakdown phenomenon can also occur in this case. However, at the same conditions, in particlular,  the same flow rates downstream of  the bottleneck and on a homogeneous road, the   nucleation rate for traffic breakdown
   on the homogeneous road   is considerably smaller than at the bottleneck~\cite{KKl2003A,KernerBook}.
   This is associated with
   empirical results in which
 the breakdown phenomenon 
 has also been observed   away from bottlenecks, however, this traffic breakdown is much more rare than at an on-ramp bottleneck~\cite{KernerBook}.
 A consideration of a nucleation model of the breakdown phenomenon for a homogeneous road is beyond the scope of this article.
 
 As usual for each first-order phase transition observed in many other systems in natural science~\cite{Gardiner},
 the  nucleation rate for traffic breakdown (\ref{probability_FS_approx}) is an exponentional function of   $\Delta_{\rm c}$ 
(\ref{overcr}). For traffic flow, in accordance with (\ref{probability_FS_approx}) and (\ref{overcr})
the exponential growth of the   nucleation rate with $\Delta_{\rm c}$ (\ref{overcr}) 
is  very sensible to the critical value for the deterministic breakdown phenomenon $q^{\rm (B)}_{\rm determ, \ FS}$.
This emphasizes the important impact of  the deterministic cluster, which occurs at the bottleneck at  $q_{\rm on}>0$,
on the     nucleation rate for traffic breakdown (\ref{probability_FS_approx}) at a given total flow rate $q_{\rm sum}$.

\section{Discussion}
\label{Sect_Dis}

 \subsection{Numerical simulations of general results of nucleation model for traffic breakdown \label{Num_Sym_S}}

Let us 
compare  general results of the nucleation model presented in Sect.~\ref{Sect_time} with the diagram of congested patterns at an on-ramp
bottleneck postulated in~\cite{Kerner2002B} and found in numerical simulations
of microscopic traffic flow models~\cite{KKl,KKW}, as well as with a microscopic theory of
the breakdown phenomenon~\cite{KKl2003A}. To reach this goal, we consider
an  example of the function $w_{-}$ (\ref{w_final})
\begin{eqnarray}
w_{-}(N)=N \bigg[ \frac{a}{1+(N/N_{0})^{4}}+b \bigg],
\label{analytical}
\end{eqnarray}
where $a$, $b$, and $N_{0}$
are functions of $q_{\rm on}$:
$a(q_{\rm on})=1.32q_{0}(q_{\rm on})/N_{0}(q_{\rm on})$ 1/h,
$q_{0}(q_{\rm on})=2700+370(1+q_{\rm on}/300)^{-1}$ vehicles/h,
$b(q_{\rm on})=33+10(1+q_{\rm on}/250)^{-1}$ 1/h,
$N_{0}(q_{\rm on})=25-6.5(1+q_{\rm on}/300)^{-1}$ vehicles;
the unit of $q_{\rm on}$ is  vehicles/h.

The analytical function (\ref{analytical})   allows us to perform a numerical 
analysis of the mean time delay (\ref{time_FS}) and the associated   nucleation rate (\ref{Probab_FS}) for the breakdown phenomenon
(F$\rightarrow$S transition).
For the analysis of (\ref{time_FS}) and (\ref{Probab_FS}), only branches $N^{\rm (determ)}$ and $N_{\rm c}$ (Fig.~\ref{q_rho})
of the function (\ref{analytical}) associated with the deterministic and critical clusters within which $N\leq N_{\rm s}$  are relevant.
This is because the maximum possible value of $N=N_{2}$ in the potential barrier $\Delta \Phi$
(\ref{delta_pi}) that determines the nucleation rate (\ref{Probab_FS}) is equal to $N_{\rm s}$.

 \begin{figure}
\begin{center}
\includegraphics[width=10 cm]{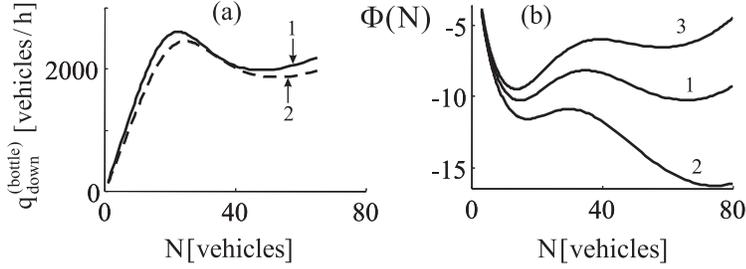}
\caption{N-shaped function $q^{\rm (bottle)}_{\rm down}(N)$ (\ref{analytical}) (a)
 for $q_{\rm on}=100$ vehicles/h (curve 1) and $q_{\rm on}=600$ vehicles/h (curve 2),
and the associated   potential $\Phi$ (\ref{potential_eq})  (b) 
as functions of the vehicle number $N$ for $q_{\rm on}=100$ vehicles/h and for three different
total flow rates $q_{\rm sum}$: 2070 (curve 3), 2200 (curve 1), 2400 vehicles/h (curve 2).  
\label{N_pot} } 
\end{center}
\end{figure}

However, for a qualitative illustration of 
a possible synchronized flow state resulting from an  F$\rightarrow$S transition, in (\ref{analytical}) the branch
for the synchronized flow state is  added, in which the detachment rate
$w_{-}(N)$ increases with $N$~\footnote{For more detail explanation of the approximation (\ref{analytical}), note that 
as in the general model (Fig.~\ref{q_rho}), the function (\ref{analytical}) is a N-shape flow--vehicle-number characteristic
(Fig.~\ref{N_pot} (a)). 
This N-shape  is chosen to satisfy those results of a microscopic three-phase traffic theory~\cite{KernerBook}
in which a Z-shaped speed--flow  characteristic for an F$\rightarrow$S transition has been found (Fig.~\ref{FS_Intr} (b, d)).
 The branch $v^{\rm (B)}_{\rm syn, \ aver}$ on this characteristic (Fig.~\ref{FS_Intr} (d))  as well as
 the associated branch $N^{\rm (syn)}$ on the N-shape flow--vehicle-number characteristic
 (Fig.~\ref{potential} (a)) are associated with a synchronized flow state, which results from the F$\rightarrow$S transition.
 The greater the density, i.e., the vehicle number $N$ within the synchronized flow state, the greater the flow rate $q^{\rm (bottle)}_{\rm down}$
(Fig.~\ref{FS_Intr} (c, e)), i.e.,  the detachment rate  $w_{-}(N)$ (\ref{w_final}). 
In real traffic flow, the growth
 of   $w_{-}(N)$  with $N$ has obviously a limit.
 This limit is related to spontaneous moving jam emergence in synchronized flow of lower speed and greater density
(i.e., greater $N$). In this case, an SP transforms into an GP, which consists of two traffic phases,
synchronized flow and wide moving jams~\cite{KernerBook}. However, these effects are beyond the scope of this article. As mentioned above,
the simple  mathematical approximation (\ref{analytical}) of the branch of   $w_{-}(N)$ for synchronized flow states that are associated with
 $N>N_{\rm s}$  can be used, because 
  at $N>N_{\rm s}$ the function $w_{-}(N)$ has no impact on results presented in the article.}. This branch corresponds to $N>N_{\rm s}$.
Respectively, this branch of the detachment rate
$w_{-}(N)$ has no influence on the analysis of (\ref{time_FS}) and (\ref{Probab_FS}). For this reason,
a simple mathematical approximation (\ref{analytical}) of the latter branch of 
$w_{-}(N)$ is chosen,
in which the detachment rate  $w_{-}(N)$ exhibits formally unlimited growth with $N$. As mentioned, this
does not impact on results discussed below.
Moreover, as follows from (\ref{Prob_stationary_1}) (see Appendix)
probability of  cluster emergence, which size $N$ is large,  is negligible. 

   \begin{figure}
\begin{center}
\includegraphics[width=10 cm]{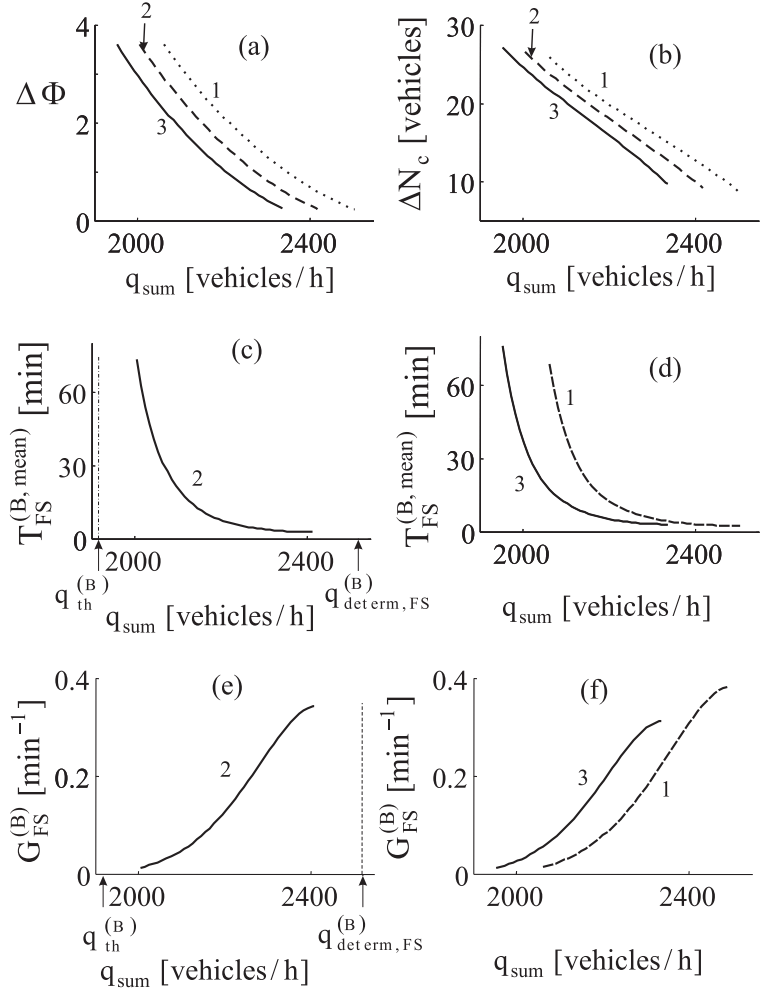}
\caption{Potential barrier $\Delta \Phi$ (\ref{delta_pi}) (a), critical vehicle number difference $\Delta N_{\rm c}$  (\ref{N_2_N_1}) (b),
mean time delay for traffic breakdown $T^{\rm (B, \ mean)}_{\rm FS}$ (\ref{time_FS}) (c, d),
and   nucleation rate for traffic breakdown $G^{\rm (B)}_{\rm FS}$ (\ref{Probab_FS}) (e, f)
as functions of the total flow rate $q_{\rm sum}$
 for three different
 flow rates $q_{\rm on}$: 100 (curves 1), 300 (curves 2), 800  (curves 3) vehicles/h.  
\label{Freq} } 
\end{center}
\end{figure}

A numerical study shows that the   potential $\Phi$ exhibits qualitatively the same   behavior at different total
flow rates $q_{\rm sum}$ (Fig.~\ref{N_pot} (b)) as those in  Fig.~\ref{potential}.
The potential barrier $\Delta \Phi$ in (\ref{time_FS}), (\ref{Probab_FS}) (Fig.~\ref{Freq} (a))
     and the  associated  critical vehicle number difference $\Delta N_{\rm c}$  (\ref{N_2_N_1}) are decreasing functions of the total flow rate
     $q_{\rm sum}$; at a given $q_{\rm sum}$ they can also be decreasing functions of  
     $q_{\rm on}$ (Fig.~\ref{Freq} (b)).
For these reasons, the total flow rate dependences of the mean time delay $T^{\rm (B, \ mean)}_{\rm FS}$ (\ref{time_FS}) 
(Fig.~\ref{Freq} (c, d))
and of the associated  nucleation rate  
 for traffic breakdown at the bottleneck (Fig.~\ref{Freq} (e, f))
  exhibit qualitative features observed in traffic flow at on-ramp bottlenecks~\cite{Persaud1998B,Lorenz2000A}
 and found in a microscopic three-phase traffic theory~\cite{KKW,KKl2003A}. 
 This confirms that the breakdown phenomenon
 at the bottleneck
 is a first-order F$\rightarrow$S transition~\cite{KernerBook}. 
 In all these curves, the  total flow rate $q_{\rm sum}$
 is smaller
 than the critical flow rate for the deterministic traffic breakdown
 $q^{\rm (B)}_{\rm determ, \ FS}$ (\ref{critical_determ}). This means that   
 $\Delta_{\rm c}$
 (\ref{overcr}) is not equal zero for all
 results  in Fig.~\ref{Freq}, i.e.,    traffic breakdown occurs
 due to a random density increase   within an initial deterministic cluster at the bottleneck.
 The total flow rate dependences of the   nucleation rate for traffic breakdown $G^{\rm (B)}_{\rm FS}$ (\ref{Probab_FS})  
  calculated at  different flow rates $q_{\rm on}$ exhibit features of three-phase traffic theory
  in which  the breakdown phenomenon
     can also occur  at  small values $q_{\rm on}$.

\begin{figure}
\begin{center}
\includegraphics[width=10 cm]{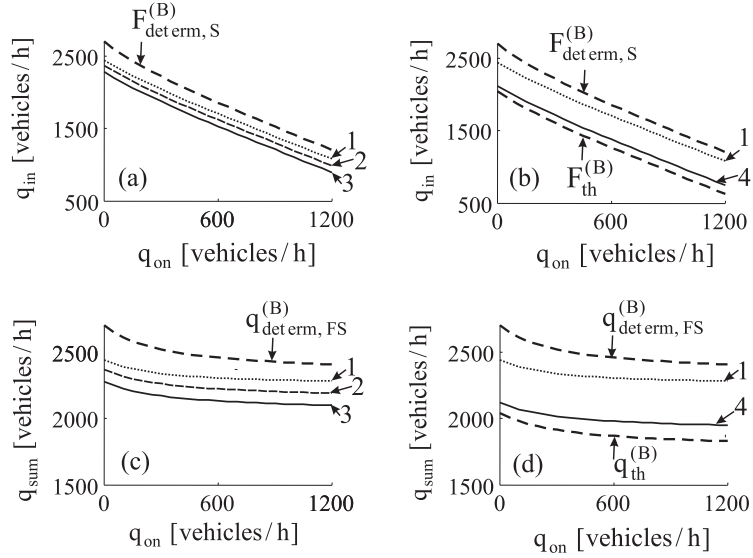}
\caption{Characteristics of the nucleation model: (a, b) -
Boundaries of constant values of the nucleation rate $G^{\rm (B)}_{\rm FS}$ of traffic breakdown (curves 1--4),
the critical boundary $F^{\rm (B)}_{\rm determ, \ S}$ for deterministic traffic breakdown (curves $F^{\rm (B)}_{\rm determ, \ S}$), and
the threshold boundary $F^{\rm (B)}_{\rm th}$ (curve $F^{\rm (B)}_{\rm th}$) as functions of the flow rates
$q_{\rm on}$ and $q_{\rm in}$. (c, d) - Dependencies of the flow rate 
 $q^{\rm (B)}_{\rm G}$  (\ref{critical_no}) (curves 1--4),
the critical flow rate $q^{\rm (B)}_{\rm determ, \ FS}$  (\ref{critical_determ})
for deterministic traffic breakdown (curves $q^{\rm (B)}_{\rm determ, \ FS}$), and
the threshold flow rate $q^{(B)}_{\rm th}$ (\ref{critical_thresh}) (curve $q^{(B)}_{\rm th}$) as functions of $q_{\rm on}$.
Curves 1--4 are related to different given values $\zeta$ for the nucleation rate of traffic breakdown $G^{\rm (B)}_{\rm FS}$  in (\ref{critical_no_p}): 
1/3.5 (curves 1), 0.2 (curves 2), 0.1 (curves 3), 1/60 (curves 4) ${\rm min}^{-1}$.  
The nucleation model cannot be applied for $q_{\rm on}= 0$, therefore, the points in all figures
in the vicinity of $q_{\rm on}= 0$ show only
the tendency of the boundaries in (a, b) and the flow rates in (c, d)
for the limiting case of small values $q_{\rm on}$  in which, however, the on-ramp inflow rate $q_{\rm on}>0$, specifically, it is assumed that the deterministic cluster
 still exists at the bottleneck.
\label{diag} } 
\end{center}
\end{figure}
    
  The critical boundary $F^{\rm (B)}_{\rm S, \ \zeta}$  (Fig.~\ref{diag} (a, b)) in the diagram of congested patterns at the
  bottleneck (flow--flow plane with the coordinates $(q_{\rm on}, \ q_{\rm in}$)) is associated with the cases in which
  the   nucleation rate for traffic breakdown is a given value $\zeta$. Therefore, the boundary $F^{\rm (B)}_{\rm S, \ \zeta}$ satisfies the condition 
  \begin{equation}
G^{\rm (B)}_{\rm FS} (q_{\rm sum}, \ q_{\rm on})=\zeta, \quad \zeta={\rm const},
 \label{critical_no_p}
 \end{equation}
i.e.,   at the boundary $F^{\rm (B)}_{\rm S, \ \zeta}$ the flow rate
 \begin{equation}
q_{\rm sum}=q^{\rm (B)}_{\rm G}(q_{\rm on})
 \label{critical_no}
 \end{equation}
depends
on $q_{\rm on}$. 
This boundary
  is qualitatively similar with the critical boundary $F^{\rm (B)}_{\rm S}$ in the diagram at which the probability for
  traffic breakdown for a given time $T_{\rm ob}$ for observing traffic flow is 1~\cite{KernerBook}.  In the diagram, there is also
  the threshold boundary $F^{\rm (B)}_{\rm th}$ (curve $F^{\rm (B)}_{\rm th}$ in Fig.~\ref{diag} (b))
  at which  the condition (\ref{critical_thresh}) is satisfied. The threshold boundary  
 also  exhibits the same qualitative features as those  found in simulation of phase transitions and spatiotemporal congested patterns
 in a microscopic three-phase traffic theory~\cite{KKl2003A,KernerBook}.
 In particular, in the limiting case of small values $q_{\rm on}$ (but $q_{\rm on}>0$,
i.e.,  it is assumed that  the deterministic cluster
 still exists at the bottleneck) the   flow rate
 $q^{\rm (B)}_{\rm G}$ reaches the maximum (limit) value $q^{\rm (B)}_{\rm G, \ lim}$ at a given   nucleation rate for 
 traffic breakdown
 $\zeta$  (\ref{critical_no_p}). 
 
 The greater the   nucleation rate for 
 traffic breakdown,
 the greater $q^{\rm (B)}_{\rm G, \ lim}$ should be. However, the increase in    nucleation rate for 
 traffic breakdown has a limit associated with deterministic traffic breakdown occurrence:  When
 $\zeta$ in (\ref{critical_no_p}) increases, the boundary $F^{\rm (B)}_{\rm S, \ \zeta}$ for random traffic breakdown   
  tends  to the 
  boundary $F^{\rm (B)}_{\rm determ, \ S}$ (curves $F^{\rm (B)}_{\rm determ, \ S}$ in Fig.~\ref{diag} (a, b))
  for   deterministic traffic breakdown in the diagram of congested patterns. 
   At the boundary  $F^{\rm (B)}_{\rm determ, \ S}$, the deterministic   breakdown phenomenon occurs within
 the deterministic cluster even if {\it no} random vehicle number increase within
 the   deterministic cluster appears at the bottleneck.
 When
 $\zeta$ in (\ref{critical_no_p}) decreases, the boundary $F^{\rm (B)}_{\rm S, \ \zeta}$     
  tends  to the threshold
  boundary $F^{\rm (B)}_{\rm th}$ (curve $F^{\rm (B)}_{\rm th}$ in Fig.~\ref{diag} (b)).
 In accordance with a microscopic   theory~\cite{KKl2003A}, in the nucleation model
  the   flow rate $q^{\rm (B)}_{\rm G}$  (\ref{critical_no}) (curves 1--4 in Fig.~\ref{diag} (c, d)),
  the critical flow rate $q^{\rm (B)}_{\rm determ, \ FS}$ for deterministic traffic breakdown (curves $q^{\rm (B)}_{\rm determ, \ FS}$ in Fig.~\ref{diag} (c, d)),
  as well as
the threshold flow rate $q^{(B)}_{\rm th}$ (\ref{critical_thresh}) (curve $q^{(B)}_{\rm th}$ in Fig.~\ref{diag} (d))
   can be the smaller, the greater $q_{\rm on}$ is.

 \subsection{Comparison with earlier nucleation models for traffic breakdown \label{Comparison_S}}
 
 In~\cite{Mahnke1997,Mahnke1999,Kuehne2002} nucleation models for traffic breakdown for a {\it homogeneous} circular road 
 have been developed (see Fig. 11 in the review~\cite{MahnkeRev}). However, rather than traffic breakdown (F$\rightarrow$S transition), in~\cite{Mahnke1997,Mahnke1999,Kuehne2002}
 a  nucleation theory for wide moving jam emergence in an initially homogeneous free  flow (F$\rightarrow$J transition)
 has been derived. Indeed, in final results of this probabilistic theory the vehicle speed within the vehicle cluster is chosen to be
 zero and a fundamental diagram for traffic flow with the vehicle cluster derived in the probabilistic theory~\cite{Kuehne2002}
  is qualitatively the same
 (see Fig. 48 in~\cite{MahnkeRev}) as those first found in~\cite{KK1994} in a macroscopic theory of free  flow metastability   associated
 with F$\rightarrow$J transition. This fundamental diagram 
 is confirmed by   empirical results associated with wide moving jam propagation (see Fig. 17 in~\cite{MahnkeRev}).
 However, even on a homogeneous road, 
 traffic breakdown is governed by an F$\rightarrow$S transition rather than by an 
 F$\rightarrow$J transition~\cite{KernerBook}. Thus, the nucleation theory of~\cite{Mahnke1997,Mahnke1999,Kuehne2002,MahnkeRev} 
 does not describe traffic breakdown on a homogeneous road.
 
 The F$\rightarrow$S transition that can occur away from freeway bottlenecks 
 is a very rare event~\cite{KernerBook}.
 This is because a freeway bottleneck introduces a spatial non-homogeneity in free flow at the road. The average speed within this
 non-homogeneity, which is permanently localized in a neighborhood of the bottleneck location, is lower and the vehicle density is greater
 than on the road away from the bottleneck~\cite{KernerBook}. This explains why in empirical observations traffic breakdowns are mostly observed at
 bottlenecks~\cite{Elefteriadou1995A,Persaud1998B,Lorenz2000A,KernerBook}.
 
The first nucleation model for traffic breakdown at an on-ramp bottleneck has been suggested by K{\"u}hne, Mahnke 
et al.~\cite{Kuehne2004,MahnkeRev}. In this model (see Chap. 17 in~\cite{MahnkeRev}),   
 a hypothesis  of three-phase traffic theory
about the sequence of the F$\rightarrow$S$\rightarrow$J transitions that governs phase transitions at the bottleneck~\cite{KernerBook}
 have been
taken into account. In addition, in accordance with this theory~\cite{KernerBook} a  random vehicle cluster, whose
occurrence can lead to an  F$\rightarrow$S transition, is localized at the bottleneck~\cite{Kuehne2004,MahnkeRev}. 

However, in this nucleation model a random vehicle precluster that emerges from fluctuations
   is necessary. This precluster, which consists of one vehicle ($n=1$), should occur
in an initial hypothetical unperturbed free  flow at bottleneck in which no vehicle cluster exists before. The precluster, which can be
associated with a random decrease in speed of one of the vehicle in a neighborhood of the bottleneck,
 foregoes  subsequent vehicle cluster evolution towards  a critical cluster (critical nuclei)
 for traffic breakdown~\cite{Kuehne2004,MahnkeRev}.
The attachment rate of precluster formation $w_{+}(0)$ is  equal to the flow rate to the on-ramp $q_{\rm on}$~\cite{Kuehne2004,MahnkeRev}: 
\begin{equation}
w_{+}(0)=q_{\rm on}.
\label{K_M}
\end{equation}  
 The attachment rate (\ref{K_M}) of  vehicle precluster formation
 does not depend on the flow rate $q_{\rm in}$ upstream of the bottleneck.  However,  in real traffic flow  both
  flow rates $q_{\rm on}$ {\it and} $q_{\rm in}$ exhibit random fluctuations.
  The formula (\ref{K_M}) should be associated with the basic model assumption
 that at $q_{\rm on}=0$ no vehicle cluster can randomly appear, specifically no traffic breakdown is possible~\cite{Kuehne2004,MahnkeRev}.
 Apparently the assumption (\ref{K_M}) leads to  the nucleation rate for an  F$\rightarrow$S transition
 at the bottleneck that is proportional to the flow rate to the on-ramp~\cite{Kuehne2004,MahnkeRev}.

Whereas for a homogeneous road the model assumption for the necessity of  precluster formation~\cite{Mahnke1997,Mahnke1999,Kuehne2002}
is physically justified, this is not the case 
 for the nucleation model at the bottleneck introduced in~\cite{Kuehne2004,MahnkeRev}. To explain this, note that
in contrast with the model of a homogeneous road, at $q_{\rm on}>0$ and $q_{\rm in}>0$
an initial state
of free flow at the bottleneck   is non-homogeneous
regardless of fluctuations. This means that even no fluctuations would occur in free flow at the bottleneck, nevertheless
free   flow is non-homogeneous in a neighborhood of the bottleneck~\cite{KernerBook}.
This is because two different flows permanent merge within the merging region of the on-ramp -- the flow with the
rate $q_{\rm on}$ and the flow with the rate $q_{\rm in}$. Vehicles merging from the on-ramp onto the main road
force vehicles on the main road to slow down. In turn, these slower moving vehicles on the main road force vehicles merging
from the on-ramp onto the main road to decrease the speed too. Thus, the speed is lower and the density is greater at the bottleneck, i.e., a local cluster appears regardless of fluctuations.
Thus, a permanent and motionless (deterministic) vehicle cluster in which speed is lower and the density is greater than away from bottleneck
exists already on the road, even if there were no fluctuations in traffic flow. For this reason, the formula (\ref{K_M})~\cite{Kuehne2004,MahnkeRev}
that assumes no vehicle cluster existence without random fluctuations in free flow at the bottleneck
is in serious conflict with empirical results and  results a microscopic three-phase traffic theory~\cite{KernerBook}.

Moreover, the master equation of this model~\cite{Kuehne2004,MahnkeRev} searches the probability for a random vehicle cluster
with $n$ vehicles. In this master equation, a vehicle cluster exists ($n>0$) only then, if the precluster has appeared.   When there were no fluctuations at the bottleneck at all, then a vehicle cluster cannot appear ($n=0$) and
 no traffic breakdown is possible in the model. In contrast, in three-phase traffic theory 
deterministic traffic breakdown is possible regardless of  fluctuations. This deterministic traffic breakdown occurs when
the size of the deterministic cluster exceeds some critical value. This is the consequence of
the non-homogeneity in free flow at the bottleneck mentioned above.

Thus, in contrast with the above basic assumptions of Ref.~\cite{Kuehne2004,MahnkeRev},  
a nucleation model that can be adequate with empirical results 
 should search  probability $p$ for random vehicle cluster evolution in which the cluster size $N$ 
 randomly changes due to fluctuations in a neighborhood of the deterministic cluster (Fig.~\ref{Cluster}).
 The size of this deterministic cluster $N^{\rm (determ)}$ does not depend on fluctuations in traffic flow.
 Random fluctuation  either increases the speed  within the cluster  or decreases it. Consequently, the density
and the cluster size $N$ either decreases or increases. In the latter case, traffic breakdown occurs at a smaller flow rate
$q_{\rm sum}=q_{\rm in}+q_{\rm on}$ than the critical flow rate $q_{\rm sum}=q^{\rm (B)}_{\rm determ \ FS}$ associated with the deterministic traffic breakdown that occurs
without any fluctuations at the bottleneck.

 These fundamental differences in the nucleation model of Ref.~\cite{Kuehne2004,MahnkeRev} and the model presented in this article
 can explain  different results of these models. 
In~\cite{Kuehne2004,MahnkeRev}, 
 even if the on-ramp inflow rate  $q_{\rm on}$ is high, in an initial steady state of traffic flow
 there is no deterministic cluster at the bottleneck. As a result,  
there is no deterministic breakdown phenomenon of three-phase traffic theory in this   model. 
Probably for this reason, in the model~\cite{Kuehne2004,MahnkeRev}
the    nucleation rate for the breakdown phenomenon 
(generation rate of traffic breakdown critical nuclei)  is proportional  to the on-ramp inflow rate
$q_{\rm on}$. As a result, if  $q_{\rm on}$ decreases below a small enough  value (but $q_{\rm on}>0$)
and the total flow rate $q_{\rm sum}$  increases (through an increase in $q_{\rm in}$),
a reasonable  given   nucleation rate for traffic breakdown  at the bottleneck
(the nucleation rate should be greater than  $\approx 1/20 \ {\rm min}^{-1}$, in accordance with empirical 
observations~\cite{Persaud1998B,Lorenz2000A})  cannot be reached.
This is true in the nucleation model of~\cite{Kuehne2004,MahnkeRev}  
  even if the total flow rate is equal to a 
  critical value  
associated with the critical nuclei for traffic breakdown consisting of {\it one vehicle} only
(in~\cite{Kuehne2004,MahnkeRev} this critical value is denoted by $q_{\rm c2}$). 
 
 In contrast,  in our model the nucleation rate (generation rate of critical nuclei)   for the breakdown phenomenon is {\it not} proportional  to $q_{\rm on}$
 and, therefore, as mentioned in Sect.~\ref{Sect_time},   for the limiting case of small values $q_{\rm on}$  (however, we assume that
   $q_{\rm on}>0$, specifically,  the deterministic cluster
 still exists at the bottleneck)
 this   generation rate of critical nuclei depends  on the total flow rate $q_{\rm sum}$ {\it only},
 i.e., this generation rate   does {\it not} depend on $q_{\rm on}$. At a given $q_{\rm sum}$, an increase in  $q_{\rm on}$
 can influence  {\it only} on such characteristics of traffic breakdown as    the critical flow rate $q^{\rm (B)}_{\rm determ, \ FS}$  
and    the threshold     flow rate $q^{\rm (B)}_{\rm th}$, as well as on  congested traffic states
at the bottleneck
 that result from the breakdown phenomenon.   
 
 When $q_{\rm sum}$ increases,
 the nucleation rate of traffic breakdown increases in the both models. 
 However,  in our model the nucleation rate cannot exceed the  nucleation rate for traffic breakdown
 associated with the deterministic breakdown
 phenomenon. In contrast with assumptions of the nucleation model of Ref.~\cite{Kuehne2004,MahnkeRev},
 in our nucleation model 
 the deterministic traffic breakdown   occurs 
 even {\it without} any random vehicle number increase within the initial  steady state of free flow
 at the bottleneck.  This is because
 if  $q_{\rm on}>0$, then in our model   
 there is a deterministic vehicle cluster 
 localized at the bottleneck, which   exists permanent at
 the bottleneck due to the on-ramp inflow.   
 In our model, random  traffic breakdown nucleation can occurs through a   random increase in vehicle number
  within  this  deterministic  cluster. 
 The mentioned qualitative differences in the nucleation model of Ref.~\cite{Kuehne2004,MahnkeRev} and our nucleation model are also responsible 
 for different dependences of the generation rate of traffic breakdown 
 on the total flow rate  in these nucleation models.

\appendix

\section{Derivation of nucleation rate}
\label{App}

In order to derive formula (\ref{time_FS}),
we use a general formula for the mean time delay $T$ of escaping from the potential well
for the master equation (\ref{Prob})~\cite{Gardiner}:
\begin{eqnarray}
\label{time_general}
T=\sum^{N_{3}}_{n=N_{1}}{ \bigg[ (w_{+}p_{\rm s}(n))^{-1}
\sum^{n}_{k=0}p_{\rm s}(k)  \bigg ]},
\end{eqnarray}
where $p_{\rm s}(N)$ is a steady solution of (\ref{Prob}), (\ref{Prob0}):
\begin{eqnarray}
p_{\rm s}(N)=p_{\rm s}(0) \prod^{N}_{n=1}\frac{w_{+}}{w_{-}(n)}  \quad \textrm{at $N>0$}.
\label{Prob_stationary}
\end{eqnarray}
When   $N_{\rm i} \gg 1, \ i=1,2$
(more rigorous conditions are given below),
the distribution $p_{\rm s}(N)$ has a sharp maximum at  $N=N_{\rm 1}$,
  and the function $p^{-1}_{\rm s}(n)$ in (\ref{time_general}) has
a sharp maximum at  $n=N_{\rm 2}$. Then the formula 
 (\ref{time_general}) can  be written as follows~\cite{Gardiner}:
\begin{eqnarray}
T=(w_{+})^{-1}\sum^{N_{2}}_{n=0}{p_{\rm s}(n) } \sum^{N_{3}}_{n=N_{1}}{ p^{-1}_{\rm s}(n)}.
\label{time_general_1}
\end{eqnarray}
Formula  (\ref{Prob_stationary}) can be written as
\begin{eqnarray}
p_{\rm s}(N)=p_{\rm s}(0) \exp{[-\Phi(N)]}  \quad \textrm{at $N \geq 0$},
\label{Prob_stationary_1}
\end{eqnarray}
where the  potential $\Phi(N)$ is given by (\ref{potential_eq}). Substituting  
(\ref{Prob_stationary_1}) into (\ref{time_general_1}),
we can find the exponentially large factor 
in  (\ref{time_general_1}) explicitly
\begin{eqnarray}
T=(w_{+})^{-1} c_{1}c_{2}\exp{[\Phi(N_{2})-\Phi(N_{1})]},
\label{time_general_2}
\end{eqnarray}
where
\begin{equation}
\label{factors}
 c_{1}=\sum^{N_{2}}_{n=0}{ \exp{[-\Delta \Phi^{(1)}(n)]}}, \
 c_{2}=\sum^{N_{3}}_{n=N_{1}}{ \exp{[\Delta \Phi^{(2)}(n)]}}, \\
\end{equation}
\begin{equation}
\Delta \Phi^{(i)}(N)=\Phi(N)-\Phi(N_{i}), \quad i=1,\ 2. 
\label{delta_exponent}
\end{equation}
The factors $ c_{1}$, $ c_{2}$ can be estimated using the parabolic approximation
of potential $\Phi(N)$ near the extremum points $N=N_{1}, \ N_{2}$~\cite{vanKampen}.
For instance, to find the factor $ c_{1}$, 
we  introduce a new variable $y=N/N_{1}$ and
approximate the sums in  (\ref{potential_eq}),  (\ref{factors})  by integrals:
\begin{eqnarray}
\Delta \Phi^{(1)}(N) \approx  \phi^{(1)}(y)=
N_{1} \int^{y}_{1} { \ln{ \frac{w_{-}(N_{1}z)} {w_{+}}}dz}, \\
 c_{1} \approx N_{1} \int^{N_{2}/N_{1}}_{0}{ \exp{[- \phi^{(1)}(y)}]}dy. 
\label{c1}
\end{eqnarray}
Using the series expansion 
\begin{equation}
\phi^{(1)}(y) = N_{1} \eta_{1}(y-1)^{2}/2+O((y-1)^{3})
\label{potential_expansion}
\end{equation}
 near the point $y=1$, where $\eta_{1}=d\ln{w_{-}}/d\ln{N}|_{N=N_{1}}$, 
we find  $ c_{1}=\sqrt{2 \pi N_{1}/ \eta_{1}}$. Similarly,  $ c_{2}=\sqrt{2 \pi N_{2}/ \eta_{2}}$, where
 $\eta_{2}=-d\ln{w_{-}}/d\ln{N}|_{N=N_{2}}$. The substitution of $ c_{1}$ and $ c_{2}$ 
into (\ref{time_general_2}) yields the formula (\ref{time_FS}).

The parabolic approximation used for estimation of
factor $ c_{1}$ holds only when we can neglect
in integral (\ref{c1}) 
third-order terms in the potential expansion (\ref{potential_expansion})~\cite{vanKampen}.
The same is true for calculation of $ c_{2}$.
The conditions of the parabolic approximation are
\begin{equation}
N_{i}\eta_{i}^{3} \gtrsim \xi_{i}^{2},  \quad i=1,2,
\label{approximation_conditions}
\end{equation}
where $\xi_{i}=N^{2}_{i}d^{2}\ln{w_{-}}/dN^{2}|_{N=N_{i}}, \ i=1,2$.

To derive the  formula (\ref{time_FS_approx_new}),
 we  approximate the value $\Delta \Phi$ in (\ref{time_FS}) by integral
\begin{eqnarray}
\Delta \Phi=N_{1} \int^{N_{2}/N_{1}}_{1} { \ln{ \frac{w_{-}(N_{1}y)} {w_{+}}}dy}.
\label{exponent}
\end{eqnarray}
Under  approximation of the function $w_{-}(N)$ in (\ref{exponent})
near the maximum point $N=N_{\rm d}$ by parabola, we get:
\begin{eqnarray}
w_{-}(N) = q^{\rm (B)}_{\rm determ, \ FS}
\left[1-\xi_{\rm d}\frac{(N-N_{\rm d})^{2}}{2N_{\rm d}^{2}}\right],
\label{w_expansion}
\end{eqnarray}
where the formula 
$q^{\rm (B)}_{\rm determ, \ FS}=w_{-}(N_{\rm d})$ is taken into account.
The roots  $N=N_{1}$ and $N=N_{2}$ of equation $q_{\rm sum}=w_{-}(N)$
given by formulae (\ref{w_final}), (\ref{steady_states_q})
are 
\begin{eqnarray}
N_{1, \ 2}=N_{d} \mp \Delta N_{\rm c}/2,
\label{roots_approx}
\end{eqnarray}
where the critical value $\Delta N_{\rm c}$ (\ref{N_2_N_1}) is
\begin{eqnarray}
\Delta N_{\rm c}=2\sqrt{2} \xi^{-1/2}_{\rm d}N_{d}\Delta_{\rm c}^{1/2}.
\label{delta_N_approx}
\end{eqnarray}
Substituting (\ref{w_expansion})--(\ref{delta_N_approx})
into (\ref{exponent}), and calculating   the derivatives
 $w^{\prime}_{-}(N_{i})$
 in (\ref{factor}):
 $w^{\prime}_{-}(N_{i}) \approx  \pm
q^{\rm (B)}_{\rm determ, \ FS}\xi_{\rm d}\Delta N_{\rm c}/(2N_{\rm d}^{2}), \ i=1,2$, we get
(\ref{time_FS_approx_new}).

Under the condition
that $q_{\rm sum}$ is close to the critical point $q^{\rm (B)}_{\rm determ, \ FS}$,
i.e., when  
\begin{equation}
\Delta_{\rm c}\ll 1,
\label{close_critical}
\end{equation}
from (\ref{time_FS_approx_new}), we get
 the approximate formula (\ref{time_FS_approx}). The latter is applicable 
under the condition (\ref{close_critical}) only if the conditions 
(\ref{approximation_conditions}) are still satisfied. Using the   formula for derivatives 
$w^{\prime}_{-}(N_{i}), \ i=1,2$ and that $N_{i}, \ i=1,2$ is
close to $N_{\rm d}$,
 we can estimate 
the values $\xi_{i} \approx \xi_{\rm d}$ 
and $\eta_{i}\approx \xi_{\rm d}\Delta N_{\rm c} /N_{\rm d}, \ i=1,2$
in (\ref{approximation_conditions}).
Thus,     the condition (\ref{approximation_conditions})
reads $\xi_{\rm d}\Delta N_{\rm c}^{3} / N^{2}_{\rm d} \gtrsim 1$.
Taking into account formula (\ref{delta_N_approx})
for $\Delta N_{\rm c}$, 
the condition (\ref{approximation_conditions}) can be written in terms of 
$\Delta_{\rm c}$: 
\begin{equation}
N_{\rm d}\Delta_{\rm c}^{3/2}\xi_{\rm d}^{-1/2} \gtrsim 1.
\end{equation}
This inequality together with (\ref{close_critical}) determine the range of
$\Delta_{\rm c}$ in which the  approximate
formula (\ref{time_FS_approx}) for the mean time delay $T^{\rm (B, \ mean)}_{\rm FS}$
of  an F$\rightarrow $S transition  at the bottleneck is valid.

\end{document}